\documentclass[showpacs,amsmath,amssymb,twocolumn,floatfix,prl]{revtex4}
\input{epsf}
\begin{document}

\title{
Stationary drag photocurrent caused by strong running wave in
quantum wire: quantization of
 current. }
\title{Stationary drag photocurrent caused by strong effective running
wave in  quantum wire: quantization of
 current}
 \author{M.V. Entin$^{(1)}$, L.I. Magarill$^{(1,2)}$}
\affiliation{$^{(1)}$Institute of Semiconductor Physics, Siberian
Branch, Russian Academy of Sciences, Novosibirsk, 630090, Russia
\\ $^{(2)}$ Novosibirsk State University, Novosibirsk, 630090, Russia}
\begin{abstract}
The stationary current induced by a strong running potential wave
in one-dimensional system is studied. Such a wave can result from
illumination of a straight  quantum wire with special grating or
spiral quantum wire by circular-polarized light.  The wave drags
electrons in the direction correlating with the direction of the
system symmetry and polarization of light. In a pure system the
wave induces minibands in the accompanied system of reference. We
study the effect in the presence of impurity scattering. The
current is an interplay between the wave drag and impurity
braking. It was found that the drag current is quantized when the
Fermi level gets into energy gaps.
\end{abstract}

\maketitle

Two main sources of the stationary photocurrent in homogeneous
systems  are known: light pressure (photon drag) \cite{grinb} and
photogalvanic \cite{belin}, \cite{ivch},\cite{we} or ratchet
effect. In the first case photons transmit their momenta to
electrons and directly accelerate electrons, in the second case
the light serves as an  energy source, while the acceleration
originates from a third body (impurities, phonons etc.), and the
current direction correlates with the polarization of light via
material tensors.

A related phenomenon is the electron drag by a surface acoustic
wave (SAW)
\cite{tal},\cite{aiz},\cite{aiz1},\cite{robin},\cite{gov},
\cite{ah},\cite{ah1},\cite{kas}. The wavelength of SAW is large as
compared with electrons, so the periodicity is less important and
electrons are treated as captured into dynamic quantum dots formed
by potential minima. The discreteness of electrons leads to the
SAW drag quantization. The quantization exists both with
 and without e-e interaction. If the wave
amplitude is weak enough the quantum dots can not keep electrons
and the picture fails.

In recent papers \cite{spiral, we2} we have studied the electron
drag by circular-polarized electromagnetic field in curved quantum
wires, particularly, in quantum spirals. In such  systems the
electric field of a long external electromagnetic
 wave is converted to an effective short wave
propagating along the wire. The wave drags electrons. The effect
resembles the travelling-wave tube with the difference that the
field remains  almost uniform while the acting component of this
field projected to the wire has a short wavelength. Besides, the
effect takes place in a solid instead of vacuum.

We have considered the problem in the limit of weak field. It was
also found that strong field bunches electrons in the potential
minima, forcing them to move with the phase speed of the wave.

It should be emphasized that an effective wave can be produced in
different ways, for example in the same way as in the
travelling-wave tube, using metallic or dielectric spiral grating
and straight quantum wire along the spiral axis.  These
inhomogeneous dielectric properties produce non-uniformity of
local electric field and form the running wave. Such a
construction permits  to use not an exotic system like
semiconductor spiral quantum wire \cite{prinz,prinz1},  but more
realistic systems: straight quantum wires together with spiral
spacial field modulators. Another more simple design is a double
grating like the one shown in Fig.1. This system also produces the
running wave near the quantum wire. Other variants of running wave
can be considered, e.g., plasmon wave.

\begin{figure}[h]\label{fig1}
\centerline{\epsfxsize=7cm \epsfbox{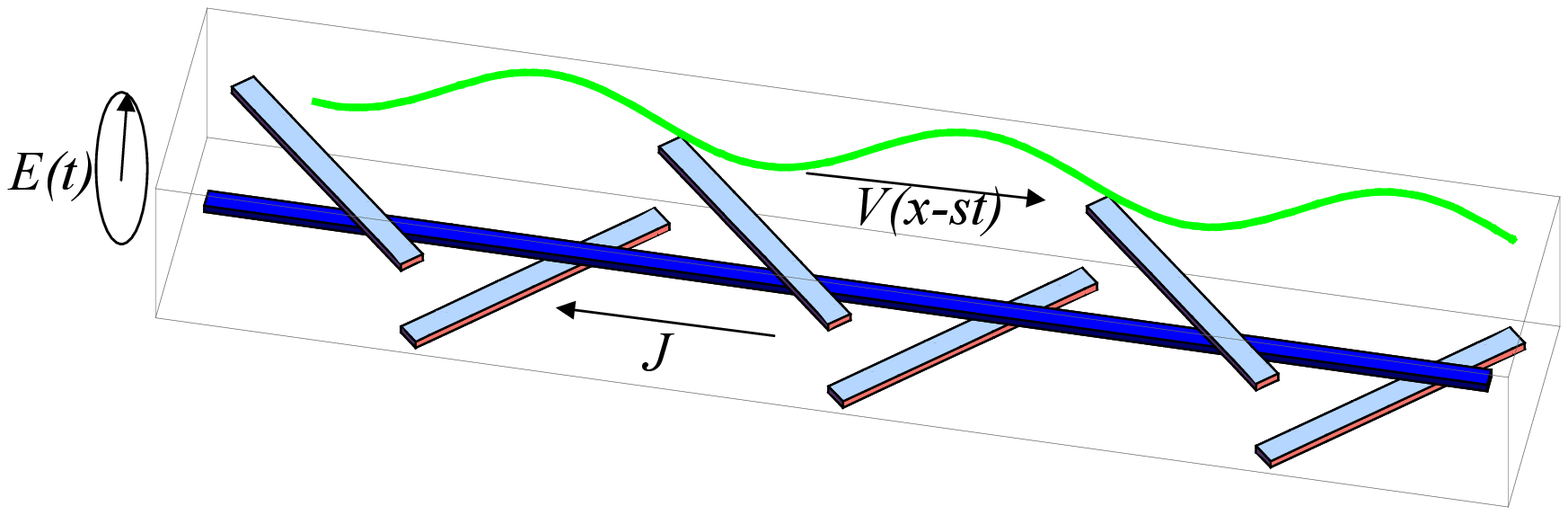} }
 \caption{(Color online) Straight quantum wire with tilted metallic (isolator) grating in  external alternating
 electric field ${\bf E}(t)$ circularly polarized in a
 plane orthogonal to the wire. The modulation of external field by the grating induces the running wave which
 drags  electrons.}
\end{figure}
The purpose of the present paper is the study of the drag current
in infinitely long homogeneous 1D system driven by the the
potential wave whose wavelength is comparable with electronic.
Such  a wave changes the electron spectrum giving rise to the
Bloch states. We have found that in these conditions the wave can
drag electrons with the velocity of the wave. This phenomenon
occurs when the Fermi level lies inside the forbidden band. That
means quantization of current $J=e\omega N/\pi$, where  $e$ is the
electron charge, $\omega$ is the frequency of the wave and $N$ is
the number of occupied bands. The accuracy of quantization is
limited by the nonlinear response on the wave velocity. We shall
demonstrate that the corrections to the quantized values are
exponentially small if the wave velocity $s$ tends to zero.

\subsection*{Basic equations}
Let us consider a strong potential wave $V=V_0\cos(kx-\omega t)$
propagating with  velocity $s=\omega/k$ along the quantum wire in
the  presence of electron scattering. The scattering caused by
impurities with the potential $U(x)=\sum_j u(x-x_j)$ is assumed.
If the field is strong one should include the wave field into the
formation of electron states and consider the impurities as a
perturbative factor. We shall use the coordinate system $x\to
x-st$ accompanying  the wave. In these coordinates the potential
of the wave is stationary and the impurities are running back with
the velocity $-s$: $U(x)\to U(x+st)$.

In the absence of impurities  electron states
$\psi_\nu(x),~~\nu\equiv (n,p)$ with a given quasimomentum $p$ in
the $n$-th band
 obey the stationary Schr\"{o}dinger equation:
\begin{equation}\label{1}
    \epsilon_\nu\psi_\nu(x)=-\frac{1}{2m}\frac{d^2}{dx^2}\psi_\nu(x)+V(x)\psi_\nu(x),
\end{equation}
and the periodicity condition $
    \psi_\nu(x+\lambda)=e^{ip\lambda}\psi_\nu(x),
$ where $\lambda=2\pi/k$ is the wavelength (we set $\hbar =1$).
The states are represented by the Mathieu functions. With regard
to the periodicity  the Mathieu functions can be written as $$
\psi_{n,p}(x)= \frac{1}{\sqrt{L}}e^{ipx}\sum_g b^n_{p+g}e^{igx},$$
$g=k r$ is the vector of reciprocal lattice, $r$ is integer, $L$
is the length of the wire. The quantities $b^n_{p+g}$ are the
Fourier harmonics of the Bloch amplitudes  satisfying  the
equations
\begin{equation}\label{Bl}
    (2m\epsilon_\nu-(p+g)^2)b^n_{p+g}-mV_0(b^n_{p+g+k}+b^n_{p+g-k})=0.
\end{equation}
The quantities $b^n_{p+g}$ are real and orthonormalized by a
condition $\sum_g b^{n'}_{p+g}b^n_{p+g}=\delta_{nn'}$.

The problem is studied in the framework of the kinetic equation
approach. The stationary  electron distribution function $f_\nu$
for electrons in the state $\nu$ obeys the kinetic equation
$
    \hat{I}_s\{f_\nu\}=0$, where the collision operator $\hat{I}_s$ includes all scattering
processes. The collision operator depends on the velocity $s$ as a
parameter. If the velocity goes to zero the impurity potential
becomes stationary and the distribution function converts to the
equilibrium Fermi function $F_\nu\equiv F(\epsilon_\nu)$. Hence,
$\hat{I}_0\{F_\nu\}=0.$

We shall assume that the phase velocity $s$ is small. In this case
$
 \hat{I}_s^{(1)}=\hat{I}_s- \hat{I}_0$ is small and one can expand
 the distribution function with respect to this smallness:
 $f_\nu=F_\nu+\chi_\nu,$
\begin{equation}\label{6}
   \hat{I}_0\{\chi_\nu\}+ \hat{I}^{(1)}\{F_\nu\}=0.
\end{equation}
The Eq. (\ref{6}) is the basic equation that determines the
corrections to the distribution function. From this point we shall
consider the impurities as  a main factor of scattering. Thus, the
collision operator can be prescribed to elastic processes caused
by impurities. The impurity collision operator reads
$\hat{I}_s\{f_\nu\}=\sum_{\nu'} W_{\nu',\nu}(f_{\nu'}-f_\nu)$

Additional simplification  with the scattering operator
$\hat{I}_s^{(1)}$ can be done by expanding it in powers of $s$. It
should be emphasized that this expansion gives a finite result if
the upper band is partially occupied (see below). The transitions
between electron states caused by moving impurities decelerate
electrons.

In the laboratory system the current is
\begin{equation}\label{7}
    j=e n_e s+e \sum_n\int_{-k/2}^{k/2}\frac{dp}{\pi}
    v_\nu\chi_\nu^{(-)},~~~~v_\nu=\frac{d\epsilon_\nu}{dp},
\end{equation}
where $\chi_\nu ^{(-)}=(\chi_\nu -\chi_{\bar{\nu}})/2$,
$\bar{\nu}\equiv (n,-p)$. The term $e n_e s$ ($n_e$ being linear
electron concentration) arises due to the transition from the
moving frame of reference to the laboratory frame.

The collision term can be expressed via scattering probability on
the moving impurities $W_{\nu,\nu'}$. In the Born approximation
the probability of scattering reads \begin{equation}\label{8}
   W_{\nu',\nu}= n_i\int dq |u(q)|^2 |J_{\nu',\nu}(q)|^2
    \delta(\epsilon_{\nu'}-\epsilon_\nu +sq),
\end{equation}
where $J_{\nu';\nu}(q)=<\nu'|e^{iqx}|\nu>$, $u(q)$ is the Fourier
transform of the potential of individual impurity and $n_i$ is the
linear density of impurities.

The expression for $\hat{I}_0\{\chi_\nu \}$ is algebraized
$$\hat{I}_0\{\chi_\nu \}=-\chi_\nu ^{(-)}/\tau_\nu,$$ where the
relaxation time is $$\tau_\nu^{-1}=\frac{n_i}{2}\int dq |u(q)|^2
\sum_{p'}|J_{\bar{\nu};\nu}(q)|^2
    \delta(\epsilon_{n,p'}-\epsilon_\nu ).$$
    The summation over $p'$  is limited by the
first Brillouin zone $|p'|<k/2$.

 The quantity $\hat{I}_s\{F_\nu
\}$ from Eq.(\ref{6}) yeilds
     \begin{eqnarray} \label{Is}
   &&\hat{I}_s\{F_\nu \}=n_i\int dq |u(q)|^2 \nonumber\times \\ &&\sum_{p'}|J_{\nu',\nu}(q)|^2
    \delta(\epsilon_{\nu'}-\epsilon_\nu +qs) [F_{\nu'}-
    F_\nu],
\end{eqnarray}

The matrix elements $J_{\nu',\nu}(q)$ can be expressed  via
$b^n_p$:
$$ J_{\nu',\nu}=\sum_g
\delta_{p'-p-q,g}B_{\nu',\nu}(g),~~~B_{\nu',\nu}(g)=\sum_{g'}b^{n'}_{p'+g'}b^n_{p+g+g'}.$$

  Thus,
\begin{eqnarray}\label{chi}\nonumber
\chi_\nu ^{(-)}=\tau_\nu n_i\int dq |u(q)|^2 \sum_{\nu' }
|J_{\nu',\nu}(q)|^2\times\\ \delta(\epsilon_{\nu' } -
    \epsilon_\nu +qs)(F_{\nu' } -F_\nu ),
\end{eqnarray} where
$ \tau_\nu ^{-1}=2n_i|v_\nu
|^{-1}\sum_g|u(2p+g)|^2|B_{\bar{\nu},\nu}(g)|^2.
$

\subsection*{Metallic case}
Expanding Eq.(\ref{chi}) by $s$ we find
\begin{eqnarray}\label{11}
    \chi_\nu^{(-)}=-\frac{s\tau_\nu}{2} \sum_g\int_{-k/2}^{k/2} dp'2\pi n_i |u(p'-p-g)
    |^2
    \nonumber\\\times (p'-p-g)|B_{\bar{\nu},\nu}(g)|^2
    \delta(\epsilon_{n,p'}-\epsilon_\nu )\frac{d }
    {d\epsilon}F(\epsilon_\nu ).
\end{eqnarray}
It is seen from Eq.(\ref{11}) that $\chi_\nu^{(-)}=0$ at zero
temperature if the Fermi level lies outside the permitted band. If
the Fermi level is inside the permitted band one can get to
\begin{eqnarray}\label{jm}\nonumber
    j=\frac{e\omega}{\pi}\Big( N+((-1)^N-1)/2+(-1)^N\times\\
    \frac{\sum_g(gd/2\pi)|u(2p_0+g)|^2|B_{N,-p_0;N,p_0}(g)|^2}{\sum_g|u(2p_0+g)|^2|B_{N,-p_0;N,p_0}(g)|^2})\Big)
    ,
\end{eqnarray}
where $N$, and $p_0>0$ satisfy the equation $\epsilon_N(p_0)=\mu$;
$N$ is the number of the last (partially) occupied permitted band,
$p_0$ is the Fermi momentum. The first term in Eq.(\ref{jm})
originates from the first term in Eq.(\ref{7}) and gives quantized
values when $\mu$ goes outside the permitted bands.

The Equation (\ref{11}) obtained in linear in $s$ approximation
 yields zero current in the accompanied system of reference if the
Fermi level gets into forbidden bands. In this case the addition
to the current, due to the transformation into the laboratory
system gives $es n_e=e\omega N/\pi$, where $N$ is the number of
the upper occupied band. Thus, the current {\it becomes
quantized}.

  Current (\ref{jm}) does not depend on the amplitude of a
scattering potential. If the amplitude of the wave goes down the
current tends to zero. This can be proved using the expression for
$|B_{N,-p_0;N,p_0}(g)|^2$ in the limit of empty lattice ($V_0=0$):
$|B_{N,-p_0;N,p_0}(g)|^2=\delta_{g,- Nk}$ if $N$ is even and
$|B_{N,-p_0;N,p_0}(g)|^2=\delta_{g,(N-1)k}$ if $N$ is odd.

 We have calculated the current according to Eq.(\ref{jm}) for two
 types of impurity potential: short range ($u(q)=const$) and the
 Coulomb potential  with $u(x)=1/\sqrt{x^2+\rho^2}$ and $u(q)= K_0(q\rho)$
 ($\rho$ is the distance from impurities to the wire being assumed  straight
 in this case, $K_n(x)$ is the modified Bessel function of the second kind).

 The results are depicted in Figs. 2. The current exhibits quantized values in the forbidden bands
 and steep decrease in the permitted band near the bands edges.  The direction of current everywhere  is opposite to the wave. That
reflects the drag of electrons induced by the wave.  Mean value of
current drops at $V_0\to 0$ or $\mu\to\infty$ due to the
perturbative character of the drag, controlled by the parameter
$(V_0/\mu)^2$. In fact, the dependence in minima corresponds to
the classical model \cite{spiral}. The large energy of electrons
results in the weakness of the wave and quasiclassical behavior of
states: in the higher permitted bands electrons almost do not
"feel" the wave.

This classical behavior is reproduced in the quantum case except
for the vicinity of the narrow gaps, where the Bragg reflection
occurs $\mu \approx \pi^2 (2r+1)^2/2 md^2$, $r=0,1,2...$.  This
reflection "pins" electron to the wave, resulting in the quantized
values of the current. In other words, full occupation of a band
blocks transitions between this and empty band at low wave
velocity. The transition between two regimes occurs in the energy
distance of the gap order. That results in very steep slopes of
the current dependence.
\begin{figure}[h]
\centerline{ \epsfxsize=6cm\epsfbox{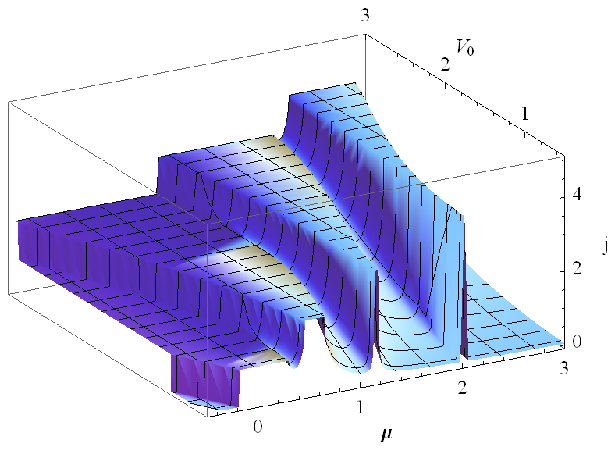}} \centerline{
\epsfxsize=6cm\epsfbox{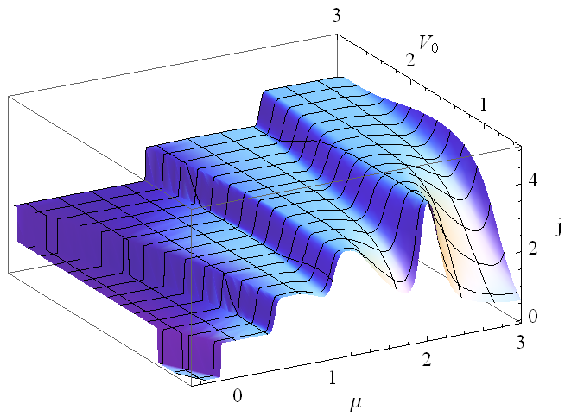}}
 \caption{ (Color online)
 Above: Drag current in units $e\omega/\pi$ {\it versus}
 the Fermi level $\mu$ and the wave amplitude $V_0$ in the case of
 short-range impurity potential $u(q)=\mbox{const}$.
 When $\mu$ gets into the forbidden bands, the current obtains integer values.
 Below: The same as in  Fig.2 for $u(q)=K_0(q\rho)$ (Coulomb impurities).}
\end{figure}
The current approaches the quantized values from below when $\mu$
approaches the edges of  permitted bands from their interior. This
is explained by the character of the drag in the wave frame,
namely, the drag of electrons near bottoms and the drag of holes
near tops.
\subsection*{Insulator case}
The previous consideration was based on the expansion with respect
to the wave speed. This expansion yields exactly quantized values
in the gaps. The corrections to the quantized values can be found
from Eqs. (\ref{chi}) without expansion on the powers of $s$. This
procedure results in the expression
\begin{eqnarray}\label{jd}\nonumber
    j=\frac{e\omega N}{\pi}-en_i\int_{-k/2}^{k/2} \frac{dp}{\pi}
    \sum_{g;p'}\sum_{n\leq N<n'}^{\infty}
    \frac{l_\nu  -l_{\nu' } }{|v_{\nu' } +s|}\times\\ |B_{\nu',\nu}(g)|^2|u(p'-p-g)|^2
\end{eqnarray}
Here $l_\nu =v_\nu \tau_\nu $; $p'$ satisfies the equation
$\epsilon_{\nu' } +sp'=
    \epsilon_\nu +(p+g)s$ (the summation over all roots is assumed).

The Pauli principle together with the conservation law permits
transitions  from occupied to empty bands only. At small $s$ the
current in the insulating state is determined by the transitions
between the last occupied and the first empty bands and $g\sim
\Delta_N/s$, where $\Delta_N$ is a gap between these bands. The
quantities $B_{N+1,p';N,p}(g)$ rapidly decay with $g$ (and, hence,
at $s\to 0$): $\log|B_{N+1,p';N,p}(g)|\propto -\Delta_N/ks$. At
the same time the other factors in Eq.(\ref{jd}) remain finite at
$s\to 0$. Hence, at dielectric gaps the corrections to the
quantized values are exponentially small.
\subsection{Discussion}
It is desirable to compare the quantum case studied here to the
classical drag effect considered earlier \cite{spiral}. In the
case of a strong classical wave, the current is simply $e n_e s$.
This value coincides with the quantum result if to express the
current via the electron concentration. At the same time this
dependence contains no  steps.

The situation recalls the quantum Hall effect where the steps in
the Hall current do not appear until the electron reservoir is
taken into consideration. The question arises: do the
impurity-induced local states in the energy gaps exist  in the
presence of a running wave? The answer is positive in the case of
the  slow wave if to replace the term "local" by "quasilocal". At
$s= 0$ any impurity induces local states in the gaps. At $s\neq 0$
the potential becomes non-stationary and the transitions from the
local to the free states appear. Nevertheless, similarly to
transitions between free states
 considered earlier, at $s\to 0$ the transition
amplitude and the widths of quasilocal states become exponentially
small. The presence of tails of quasilocal states at the gap
determines the reservoir and possibility of a continuous motion of
the Fermi level in the energy gaps with  electron density.

The current quantization is an allied problem to the charge
quantization in adiabatic quantum pumps \cite{tau}, \cite{brouw}.
In fact, the wave transmits exactly two electrons per  cycle of
field per an occupied band. The adiabaticity in the case
considered here is provided by the low frequency. Nevertheless,
the problems are different, since in the theory of the adiabatic
quantum pumps, the discrete spectrum is supposed, while the system
with a wave possesses a continuous spectrum.

It should be emphasized that the present approach differs from the
studies of quantized SAW drag \cite{ah},\cite{ah1},\cite{kas} by
the short length of wave resulting in the formation of the Bloch
states instead of the local states in the wave minima and
infinitely long quantum wire that demands taking the scattering
into account. The difference from \cite{aiz1},\cite{robin}  is the
absence of the e-e interaction. As a result of the spin degeneracy
the steps in the current are observed at $e\omega N/\pi$ values
instead of $e\omega N/2\pi$ and
 the current between steps (in metallic regime) goes through
minima. The current quantization is explained by the Bragg
scattering of electrons rather than the discreteness of electrons
in the scenario of moving quantum dots utilized  in the theory of
the quantized SAW drag.

\subsection*{Acknowledgments} The work was supported by grant of RFBR
08-02-00506 and Programs of Russian Academy of Sciences.

\end{document}